\documentclass{JHEP3}

\pdfoutput=1

\usepackage{graphicx}
\usepackage{amsmath,amssymb}
\usepackage{feynmf}

\newcommand{\bea}{\begin{eqnarray}}
\newcommand{\eea}{\end{eqnarray}}
\newcommand{\beq}{\begin{equation}}
\newcommand{\eeq}{\end{equation}}

\newcommand{\mpl}{m_{\mbox {\tiny Pl}}}

\title{On Inflation with Non-minimal Coupling}

\author{
Mark P.  Hertzberg
\\
~\\
Center for Theoretical Physics and Department of Physics,\\ 
Massachusetts Institute of Technology, Cambridge, MA 02139, USA\\
~\\
Kavli Institute for Particle Astrophysics and Cosmology,\\
Stanford University, Menlo Park, CA 94025, USA\\
~\\
Stanford Institute for Theoretical Physics,\\
Stanford University, Stanford, CA 94305, USA\\
~\\
\emph{Email}: \email{mphertz@mit.edu}
}

\abstract{
A simple realization of inflation consists of adding the following operators to the Einstein-Hilbert action: $(\partial\phi)^2$, $\lambda\phi^4$, and $\xi\phi^2\mathcal{R}$, with $\xi$ a large non-minimal coupling. Recently there has been much discussion as to whether such theories make sense quantum mechanically and if the inflaton $\phi$ can also be the Standard Model Higgs.
In this work we answer these questions.
Firstly, for a single scalar $\phi$, we show that the quantum field theory is well behaved in the pure gravity and kinetic sectors,
since the quantum generated corrections  are small. However, the theory likely breaks down at $\sim\mpl/\xi$ due to scattering provided by the self-interacting potential $\lambda\phi^4$.
Secondly, we show that the theory changes for multiple scalars $\vec\phi$ with non-minimal coupling $\xi\vec\phi\cdot\vec\phi\,\mathcal{R}$, since this introduces qualitatively new interactions which manifestly generate large quantum corrections even in the gravity and kinetic sectors, spoiling the theory for energies $\gtrsim \mpl/\xi$. 
Since the Higgs doublet of the Standard Model includes the Higgs boson and 3 Goldstone bosons, it falls into the latter category and therefore its validity is manifestly spoiled.
We show that these conclusions hold in both the Jordan and Einstein frames and
describe an intuitive analogy in the form of the pion Lagrangian. 
We also examine the recent claim that curvature-squared inflation models fail quantum mechanically.
Our work appears to go beyond the recent discussions.
}

\begin{document}


\section{Introduction} \label{sec:introduction}

Cosmological inflation is our leading theory of the very early universe \cite{Guth,Linde,AlbrechtSteinhardt82,Linde83}, although its underlying microphysics is still unknown.
Slow-roll inflation occurs in many models constructed over the years, with a scattering of model-dependent predictions.
Models of inflation are quite UV sensitive since it may have occurred at extremely high energy scales, far higher than that which we can probe at colliders, and since some models involve super-Planckian excursions in field space.
 This suggests that top-down approaches may be required to make progress, although progress in that direction has not been easy (e.g., see \cite{Baumann:2007np}). On the other hand, it is interesting to explore simple models to see what we might learn and if we can connect inflation to low energy physics.

One approach is to focus on dimension 4 Lagrangians,
which allows the inclusion of the operators $(\partial\phi)^2$, $\lambda\phi^4$, and $\xi\phi^2\mathcal{R}$ in addition to the Einstein Hilbert term $\mpl^2\mathcal{R}$ (by ``dimension 4" we mean that each operator has a dimensionless coefficient, although gravity modifies the power counting since every term is an infinite tower of operators if we expand around flat space). The $\xi\phi^2\mathcal{R}$ term is in fact required to exist for an interacting scalar field in curved space (although not for a Goldstone boson, for example).
By taking the non-minimal coupling $\xi$ to be large, a phase of inflation takes place, as originally discussed in \cite{Salopek} with constraints discussed in \cite{Fakir,Kaiser,Komatsu,Nozari:2007eq}. 

A large dimensionless coupling is unusual from the perspective of particle physics, leading to much recent debate as to whether such models make sense quantum mechanically \cite{Salopek,Bezrukov,Barvinsky,Bezrukov2,Bellido,DeSimone:2008ei,Bez2008,Burgess,Espinosa,Bezrukov:2009db,Barvinsky:2009fy,Clark:2009dc,Lerner:2009xg,Barvinsky:2009ii,Figueroa:2009jw,Okada:2009wz,Einhorn:2009bh,Lerner:2009na,Mazumdar:2010sa}. In this work we show that for a single scalar field $\phi$ the model is well behaved quantum mechanically in the pure gravity and kinetic sectors, since these do not generate large quantum corrections. However, the self-interacting potential does likely cause the theory to fail. For a vector of fields $\vec\phi$ carrying an $O(N)$ symmetry the theory manifestly generates large quantum corrections in the gravity and kinetic sectors and breaks down at high scales relevant to inflation. As far as we aware, this  difference between the single field and 
multi-field case has not been fully appreciated in the literature.
So multi-field models of non-minimal inflation are manifestly ruled out (including the Standard Model Higgs),
while single field models of non-minimal inflation are also problematic, in the sense that the Einstein frame potential is non-polynomial and therefore these models likely fail due to high energy scattering processes. This presents a challenge to them having a UV completion. We also examine curvature-squared models and show that their scalar-tensor formulation is perturbatively well behaved in the gravity, kinetic, and potential sectors; this contradicts the claims of Ref.~\cite{Burgess}. 

Our paper is organized as follows: in Section \ref{sec:nonmin} we briefly review the class of non-minimal models, in Section 
\ref{Quant} we discuss the quantum corrections, in Section \ref{Pion} we discuss the pion analogy, 
in Section \ref{Curvature} we discuss curvature-squared models, and finally we conclude in Section \ref{Conclusion}.

\section{Non-minimally Coupled Models}
\label{sec:nonmin}

Consider a vector of scalar fields $\vec{\phi}$ with an $O(N)$ symmetry, non-minimally coupled to gravity. 
It is permissible to introduce the following terms into the action: $\partial\vec\phi\cdot\partial\vec\phi$, $\lambda(\vec\phi\cdot\vec\phi)^2$, and $\xi\vec\phi\cdot\vec\phi\mathcal{R}$, where the last term represents a non-minimal coupling of the field to gravity. At the classical level we can choose to simply study only these terms, although all terms consistent with symmetries will be generated in the quantum theory; an issue we will turn to in the next Section on quantum corrections.
For now, adding these 3 terms to the usual Einstein-Hilbert term, we have the action
\beq
S=\!\int d^4 x \sqrt{-g}\left[\frac12 \mpl^2 f(\vec\phi) \mathcal{R} -\frac12 \partial \vec{\phi}\cdot\partial\vec\phi-V(\vec\phi) \right],
\label{actionJ}
\eeq
where $f(\vec\phi)=1+\xi\vec\phi\cdot\vec\phi/\mpl^2$ and $V(\vec\phi)={\lambda\over4}(\vec\phi\cdot\vec\phi)^2$. 
Since we have a new parameter $\xi$, which is essentially unconstrained by observation, this allows
the self-coupling parameter to be non-negligible, such as $\lambda=\mathcal{O}(10^{-1})$, and yet we can achieve the correct amplitude of density fluctuations by making $\xi$ very large: $\xi=\mathcal{O}(10^4)$. This was the original motivation of Ref.~\cite{Salopek} for introducing such models, since it appeared that embedding the inflaton into the matter sector of the theory would be easier. By contrast compare to minimal $\lambda\phi^4$ inflation, with 
$\lambda=\mathcal{O}(10^{-12})$, which is evidently radiatively unstable if $\phi$ carries appreciable couplings to other fields.

One may either study the theory in this original Jordan frame, or switch to the Einstein frame by
defining a new metric $g_{\mu\nu}^E=f(\vec\phi)g_{\mu\nu}$. The corresponding Einstein frame potential is
\beq
V_E(\vec\phi)={V(\vec\phi)\over f(\vec\phi)^2} = {{\lambda\over 4}(\vec\phi\cdot\vec\phi)^2\over 
\left(1+{\xi\vec\phi\cdot\vec\phi\over \mpl^2}\right)^2}.
\label{VEin}\eeq
In the Einstein frame, although the gravity sector is simple, the kinetic sector for $\vec\phi$ is not. Ignoring total derivatives, we find the Einstein frame action to be
\beq
S=\!\int d^4 x \sqrt{-g_E}\left[\frac12 \mpl^2 \mathcal{R}_E
-\frac12 \frac{1}{f(\vec\phi)}\partial\vec\phi\cdot\partial\vec\phi
-\frac{3\,\xi^2}{\mpl^2 f(\vec\phi)^2}(\vec\phi\cdot\partial\vec\phi)^2 - V_E(\vec\phi)\right].
\label{actionE}
\eeq
The first term of the kinetic sector comes from the simple conformal rescaling and the second term comes from transforming the Ricci scalar.

In the classical dynamics we can assume that only a single component of the $\vec\phi$ vector is rolling during inflation since all transverse modes will be frozen due to Hubble friction. Assuming only one component is active, call it $\phi$, the kinetic energy sector can be made canonical (this will change in the quantum analysis, where all components will be important). Let us make this explicit. We introduce a new field $\sigma$, defined through the integral
\beq
\sigma\equiv
\mbox{sign}(\phi)\int_0^\phi d\bar\phi \sqrt{{1\over f(\bar\phi)}+{6\,\xi^2\bar\phi^2\over 
\mpl^2f(\bar\phi)^2}}\,\label{kinetic}.
\eeq
In terms of $\sigma$ both the gravity and kinetic sectors are minimal, but the potential $V_E(\sigma)$ is somewhat complicated, but monotonic and well behaved.

It is simple to see from eq.~(\ref{VEin}) that when $\phi\gg \mpl/\sqrt{\xi}$ (corresponding to $\sigma\gg\mpl$) the potential $V_E$ approaches a constant; this is the regime in which the effective Planck mass runs in the original Jordan frame. In this regime the flatness of the potential ensures that a phase of slow-roll inflation takes place. 
Inflation ends when $\phi\sim\phi_{e}\equiv\mpl/\sqrt\xi$. For details on the inflationary predictions we point the reader to
Refs.~\cite{Salopek, Bezrukov}.

\section{Quantum Corrections}\label{Quant}

Various attempts have been made at addressing whether quantum corrections spoil the validity of these non-minimal models of inflation due to the $\xi\phi^2\mathcal{R}$ term; see \cite{Salopek,Bezrukov,Barvinsky,Bezrukov2,Bellido,DeSimone:2008ei,Bez2008,Burgess,Espinosa,Bezrukov:2009db,Barvinsky:2009fy,Clark:2009dc,Lerner:2009xg,Barvinsky:2009ii,Figueroa:2009jw,Okada:2009wz,Einhorn:2009bh,Lerner:2009na,Mazumdar:2010sa}. 
In each of these works, the conclusion (either positive or negative) came independent of the number of scalars $\phi^i$. Here we will show that the answer is in fact two-fold: 
for single field models the gravity and kinetic sectors do not generate large quantum corrections, 
instead it is the potential $V(\phi)$ that causes the theory to enter a regime in which standard perturbation theory likely breaks down. For multiple field models, we show that the theory manifestly generates large quantum corrections in the gravity and kinetic sectors and breaks down in the regime relevant for inflation.

\subsection{Single Field}\label{Single}

Most descriptions of non-minimal models have focussed on just a single field $\phi$, even for cases where $\phi$ represents the Higgs. For example, in \cite{Espinosa} it is explicitly assumed that all quantum issues require only a single field analysis (and it is concluded in \cite{Espinosa} that the quantum field theory is highly unnatural and breaks down).
In this case it is subtle as to whether the effective field theory makes sense. Let us go through the analysis first in the Jordan frame and then the Einstein frame. We begin by examining the gravity and kinetic sectors (as was the focus of Ref.~\cite{Burgess}) before addressing the potential $V$ in the next section.

\begin{figure}[t]
\begin{center}
\includegraphics[scale=0.3,angle=0]{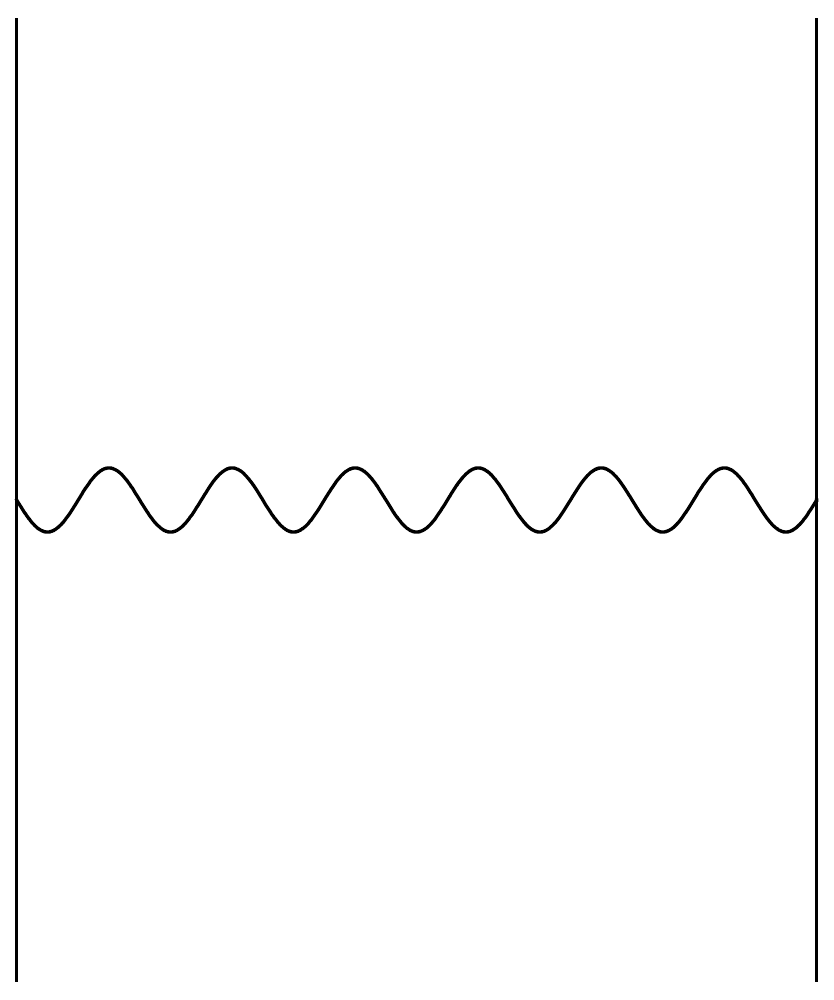}\,\,\,\,\,\,\,\,
\includegraphics[scale=0.3,angle=0]{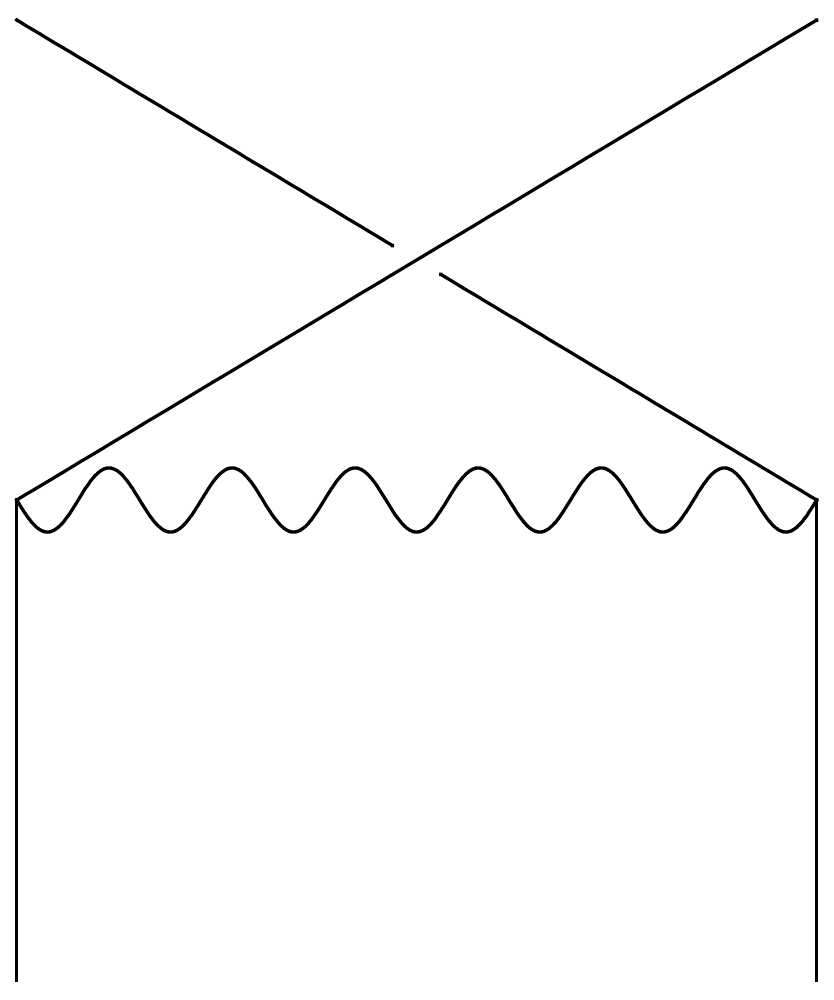}\,\,\,\,\,\,\,\,
\includegraphics[scale=0.3,angle=0]{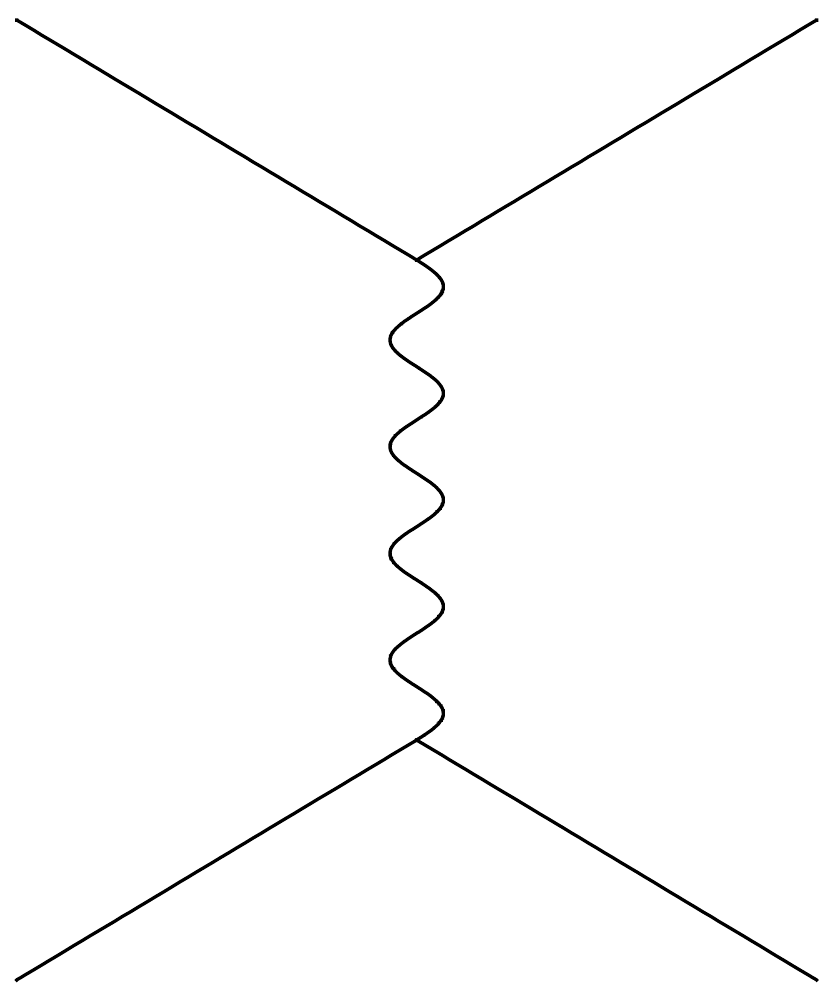}\,\,\,\,\,\,\,\,\,\,\,\,\,\,\,\,\,\,\,\,\,\,\,\,\,
\includegraphics[scale=0.3,angle=0]{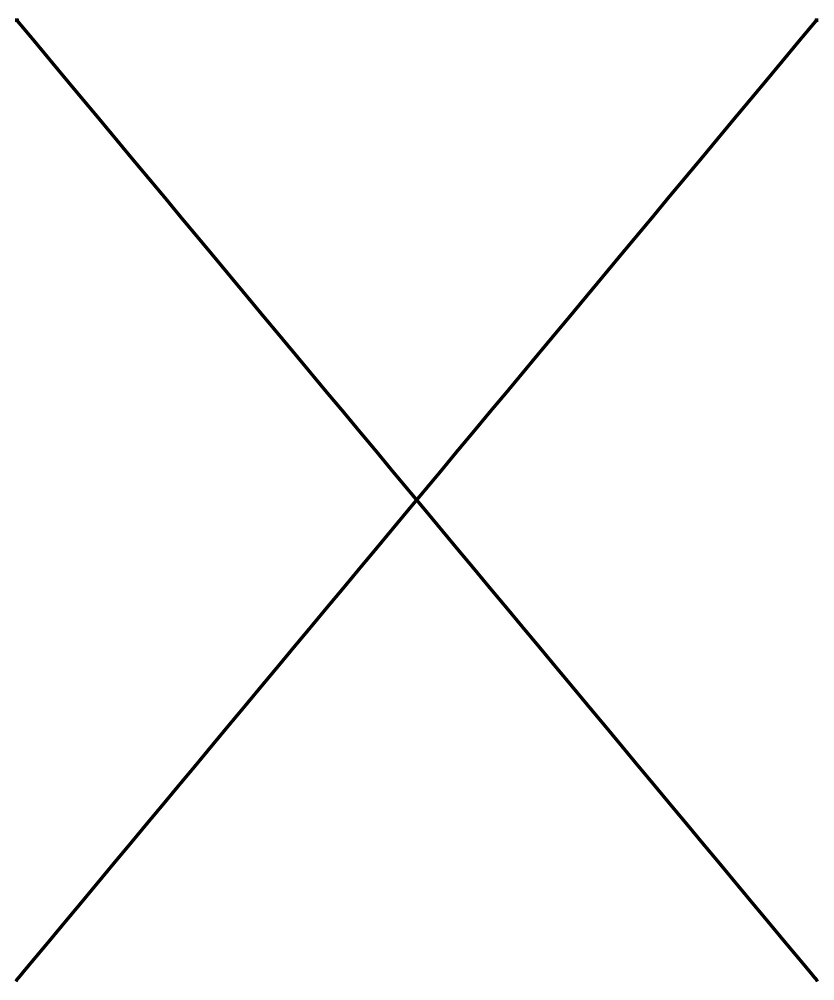}\\
\end{center}
\caption{Tree-level scattering $2\phi\to2\phi$. Diagrams 1--3: In Jordan frame it is due to graviton exchange through t, u, and s-channels. Diagram 4: In Einstein frame it is due to a single 4-point vertex.}
\label{Scattering}
\end{figure}

In the Jordan frame we must examine the consequences of the $\xi\phi^2\mathcal{R}$ operator. Let us expand around flat space. We decompose the metric as 
\beq
g_{\mu\nu}=\eta_{\mu\nu}+{h_{\mu\nu}\over\mpl}, 
\eeq
here $h_{\mu\nu}$ are metric perturbations with mass dimension 1. The Ricci scalar has an expansion around flat space that goes as $\mathcal{R}\sim\square h/\mpl +\ldots$. To leading order in the Planck mass, this gives the following dimension 5 operator in the Jordan frame action
\beq
{\xi\over\mpl}\phi^2\square h.
\eeq
It is tempting to declare that this implies the theory has a cutoff at $\Lambda=\mpl/\xi$, but is this correct?
To answer this, let us consider the scattering process: $2\phi\to2\phi$. At tree-level this proceeds via a single exchange of a graviton, giving the following leading order contribution to the matrix element (we assume massless $\phi$ particles)
\beq
\mathcal{M}_{\mbox{\tiny{c}}}(2\phi\to2\phi)\sim{\xi^2 E^2\over\mpl^2}.
\eeq
This appears to confirm that the cutoff is indeed $\Lambda=\mpl/\xi$, but this conclusion is premature. Let us explain why.
The index c on $\mathcal{M}$ stands for ``channel"; there are s, t, and u-channels, which all scale similarly (Fig.~\ref{Scattering} diagrams 1--3). When we sum over all 3 channels to get $\mathcal{M}_{\mbox{\tiny{tot}}}$ and put the external particles on-shell, an amusing thing happens: they cancel. So the leading term in powers of $\xi$ vanishes \cite{Huggins1,Huggins2}. 
The first non-zero piece is 
\beq
\mathcal{M}_{\mbox{\tiny{tot}}}(2\phi\to2\phi)\sim{E^2\over\mpl^2},
\eeq
which shows that the true cutoff  in the pure gravity and kinetic sectors is $\Lambda=\mpl$ (although the inclusion of the potential $V$ will greatly change this, as we explain shortly). Hence, according to $2\to2$ tree-level scattering the theory only fails at Planckian energies.

\begin{figure}[t]
\begin{center}
\includegraphics[scale=0.3,angle=0]{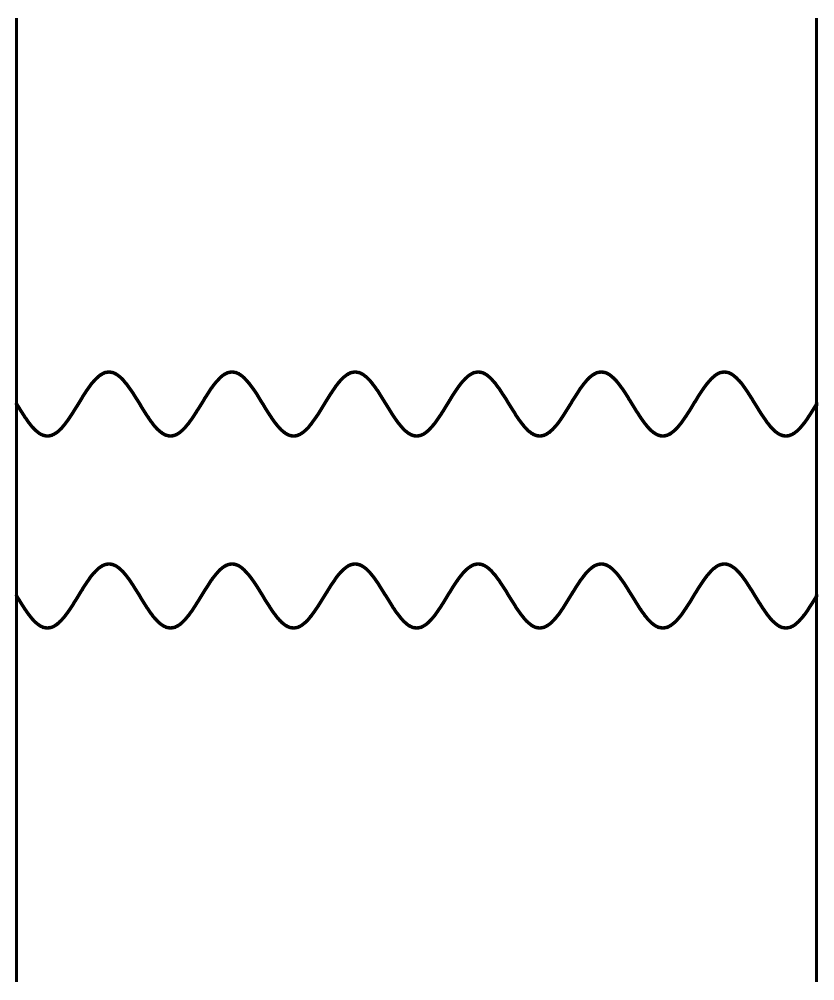}\,\,\,\,\,\,\,\,
\includegraphics[scale=0.3,angle=0]{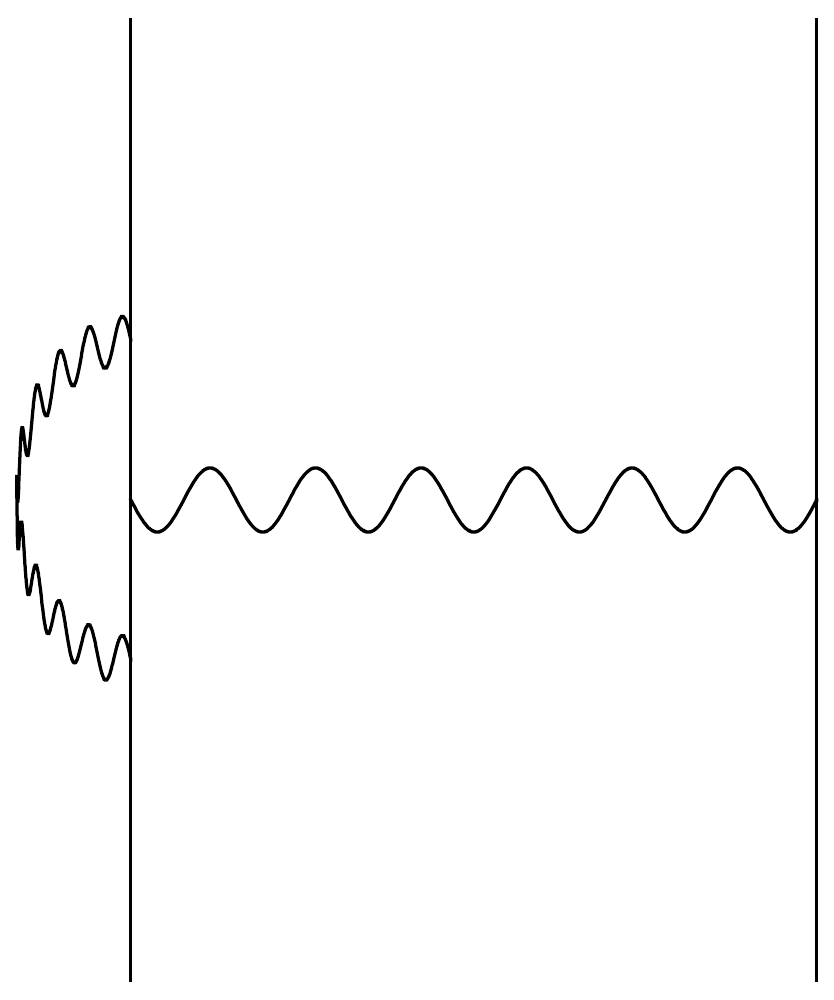}\,\,\,\,\,\,\,\,
\includegraphics[scale=0.3,angle=0]{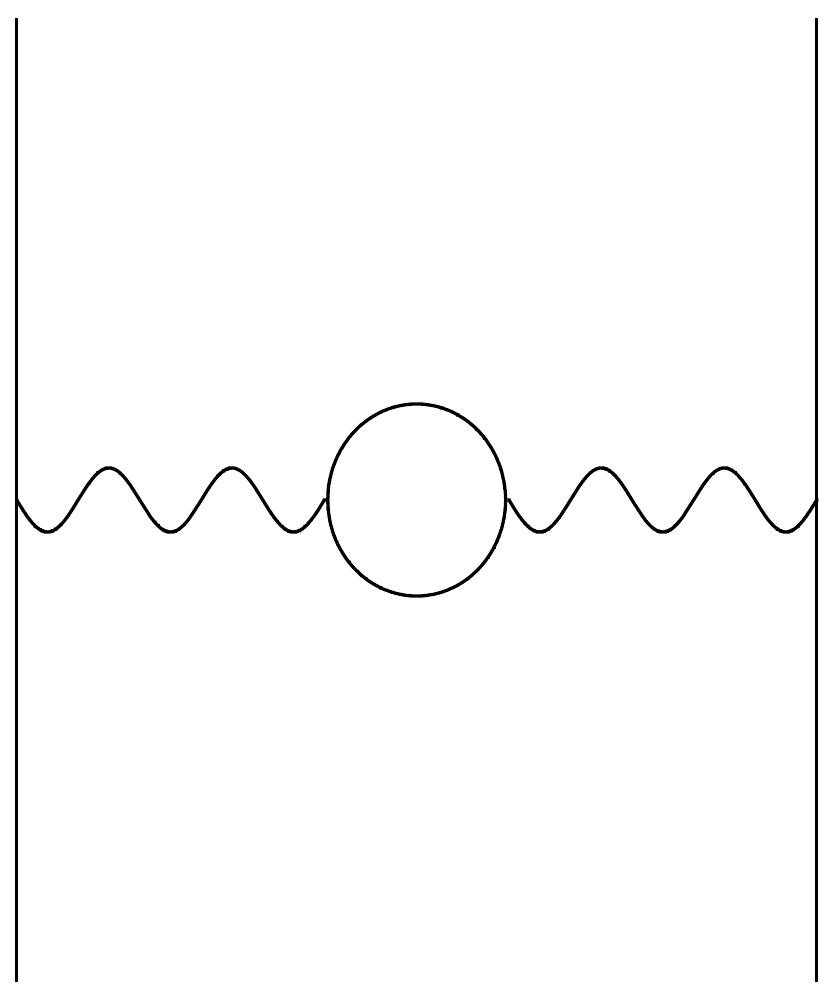}\,\,\,\,\,\,\,\,  \,\,\,\,\,\,\,\,\,\,\,\,\,\,\,\,\,
\includegraphics[scale=0.3,angle=0]{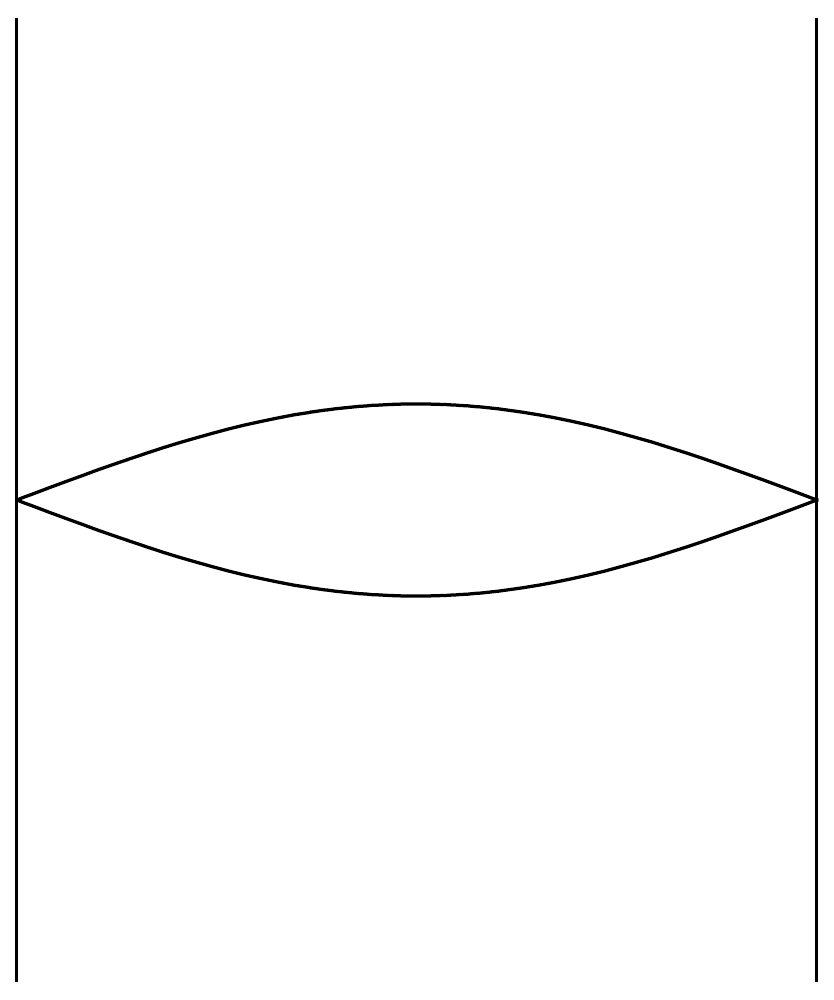}
\\ \vspace{4mm}
\includegraphics[scale=0.3,angle=0]{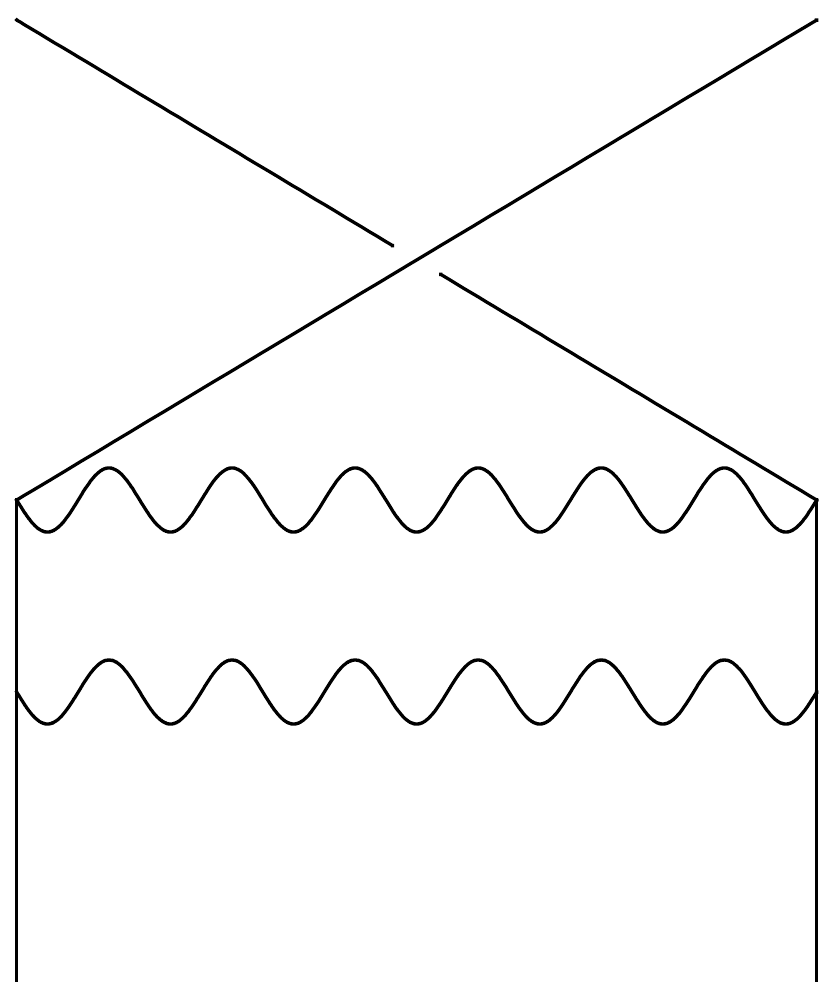}\,\,\,\,\,\,\,\,
\includegraphics[scale=0.3,angle=0]{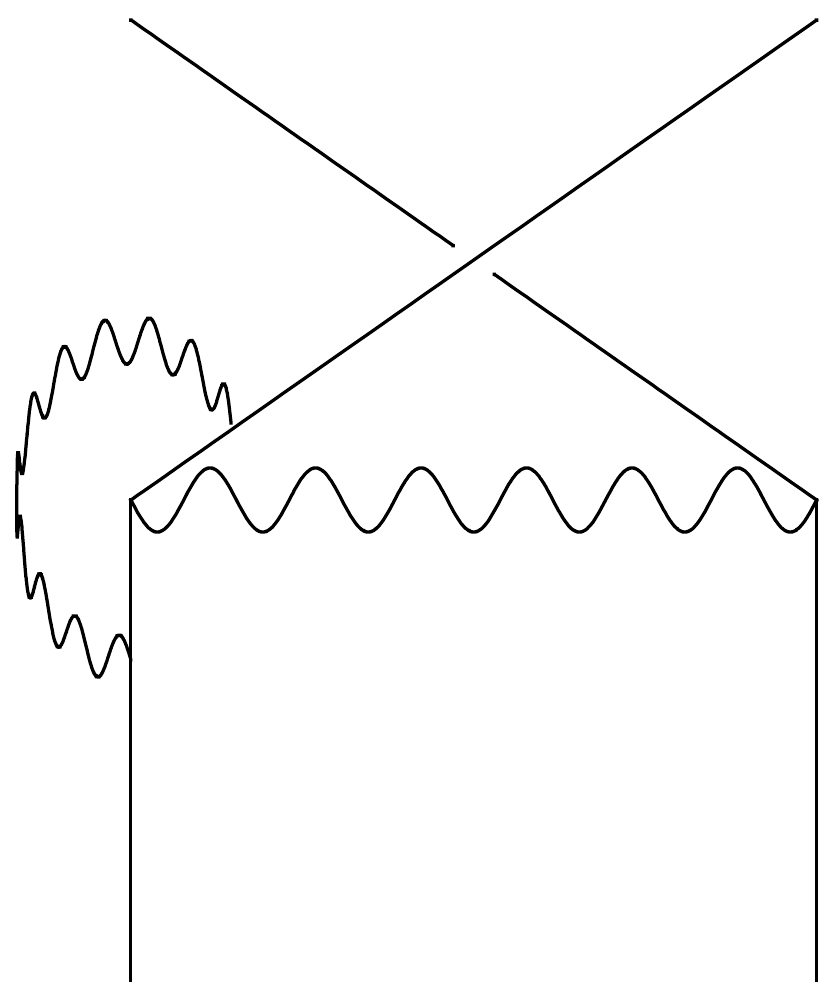}\,\,\,\,\,\,\,\,
\includegraphics[scale=0.3,angle=0]{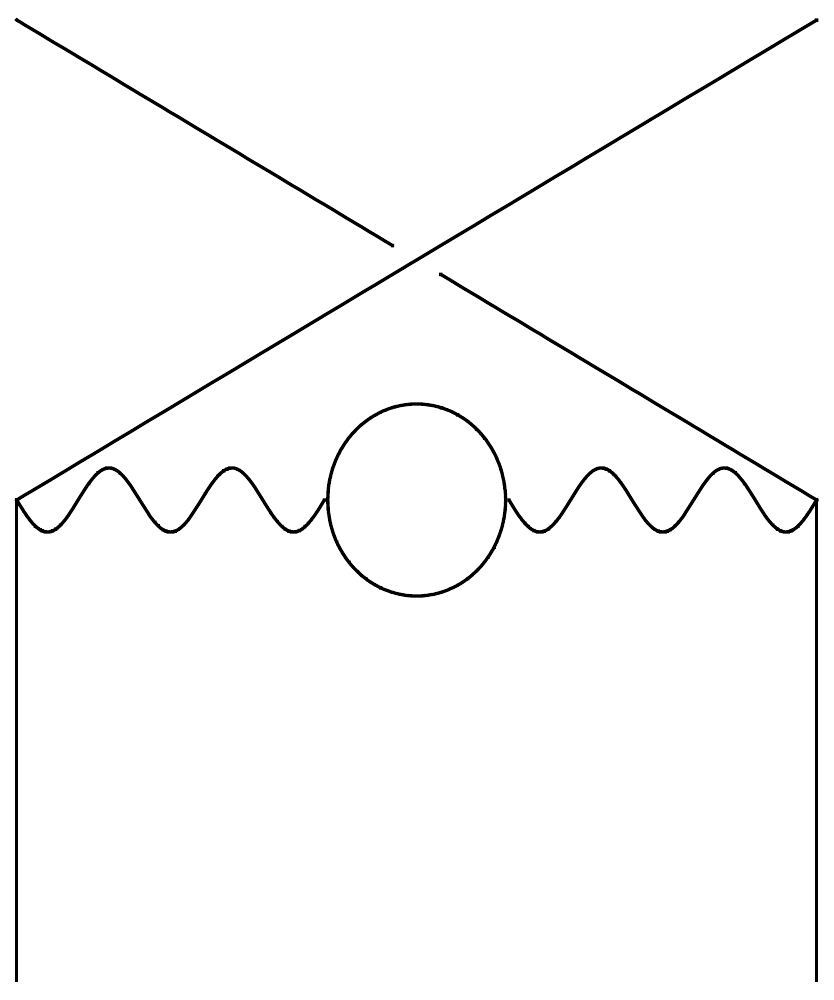}\,\,\,\,\,\,\,\,   \,\,\,\,\,\,\,\,\,\,\,\,\,\,\,\,\,
\includegraphics[scale=0.3,angle=0]{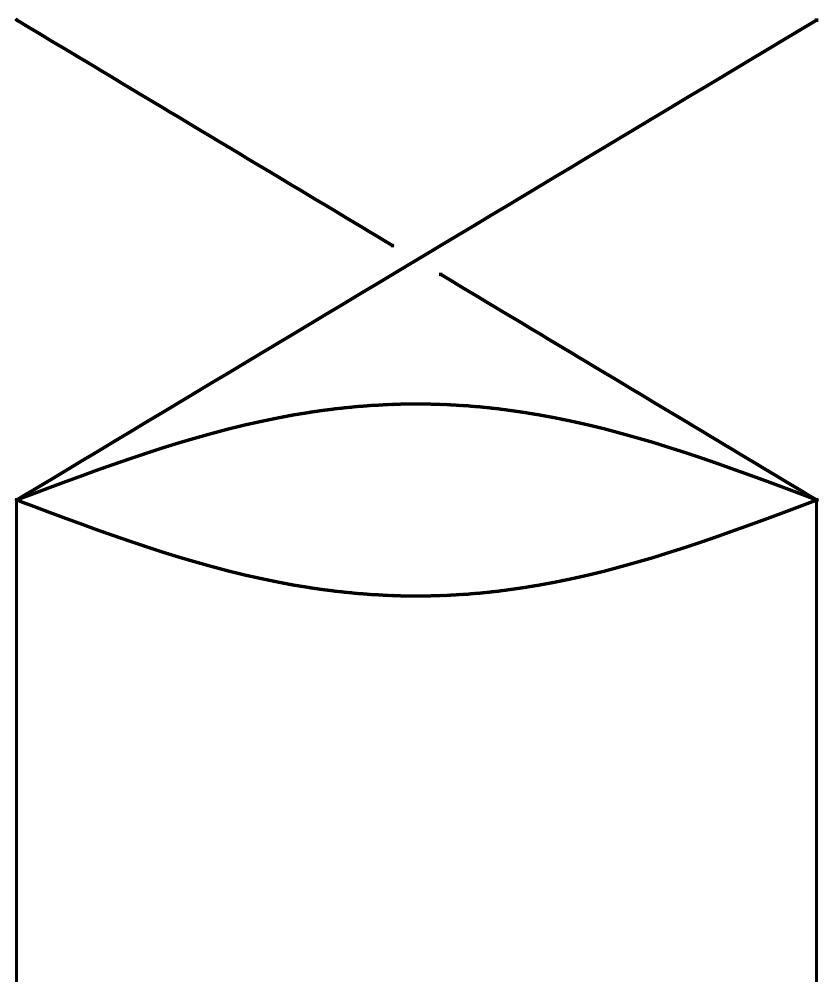}
\\ \vspace{4mm}
\includegraphics[scale=0.3,angle=0]{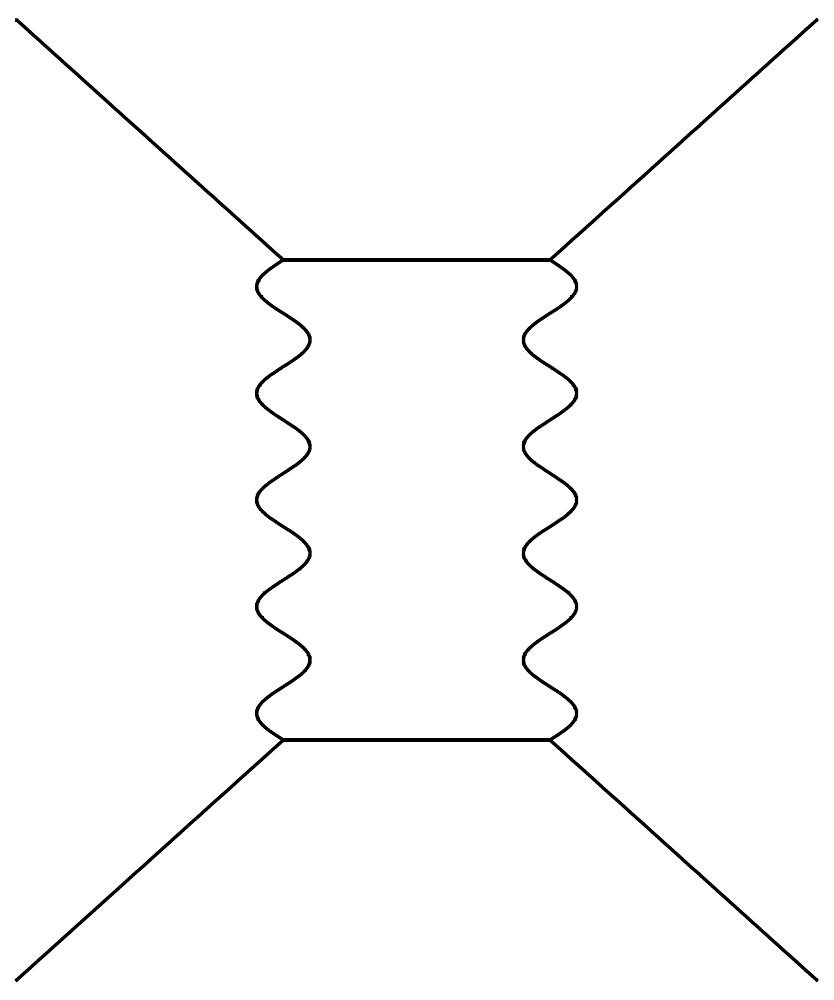}\,\,\,\,\,\,\,\,
\includegraphics[scale=0.3,angle=0]{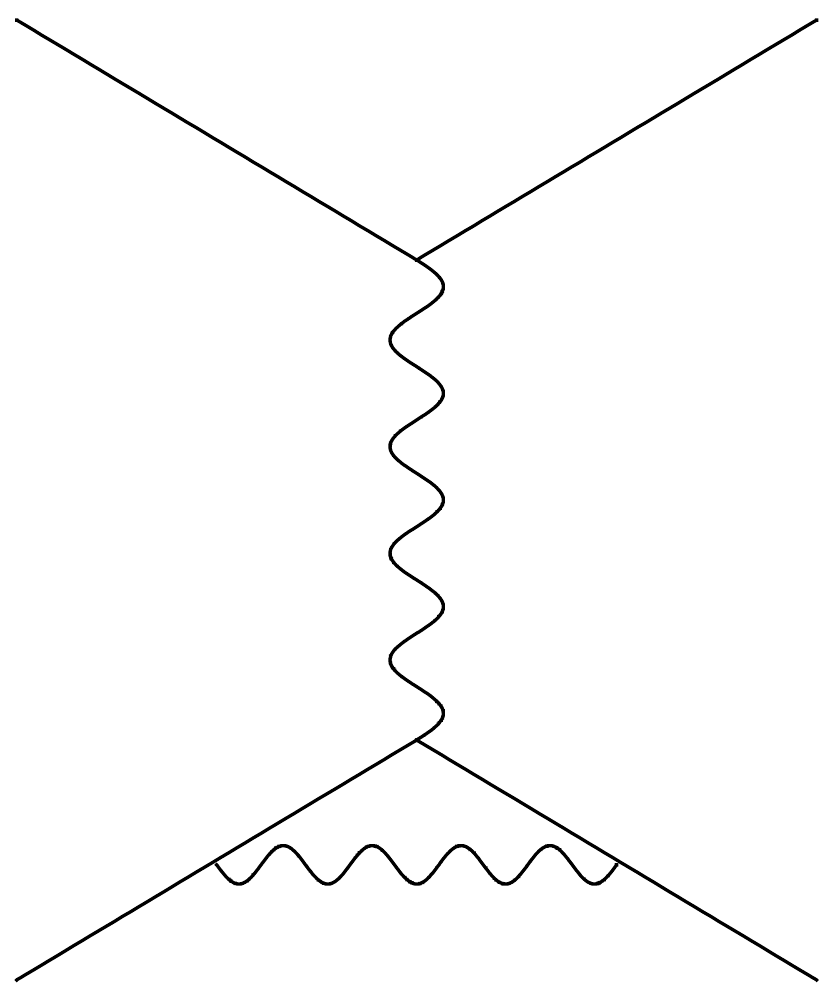}\,\,\,\,\,\,\,\,
\includegraphics[scale=0.3,angle=0]{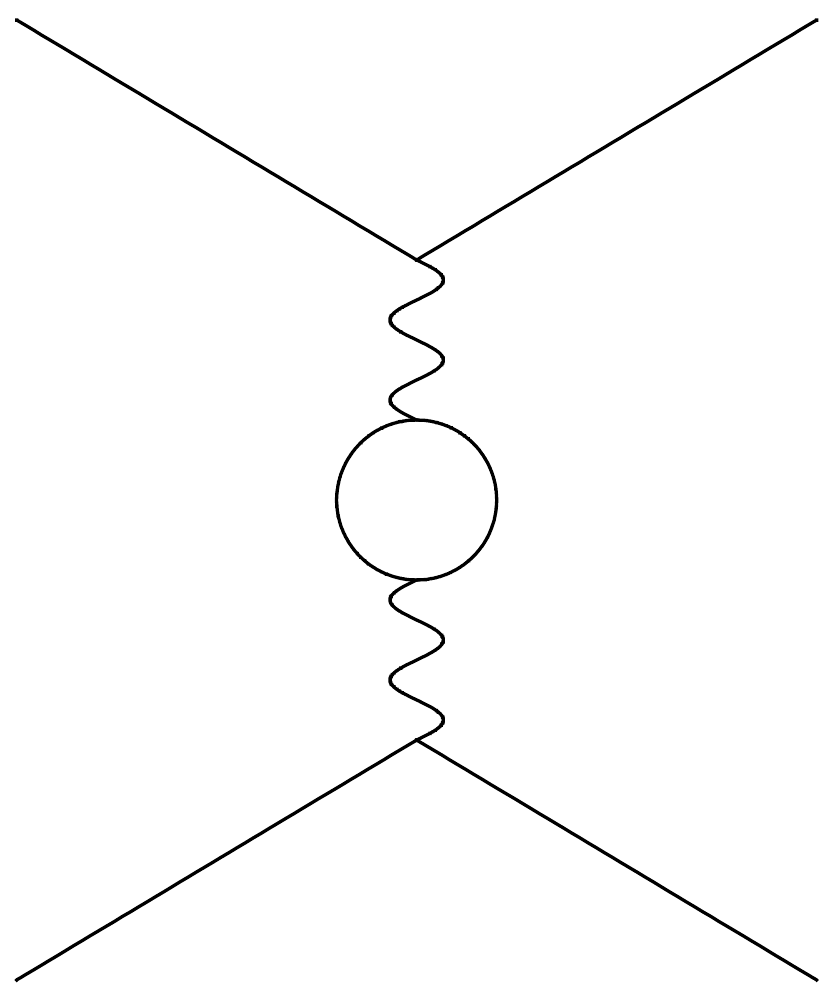}\,\,\,\,\,\,\,\,    \,\,\,\,\,\,\,\,\,\,\,\,\,\,\,\,\,
\includegraphics[scale=0.3,angle=0]{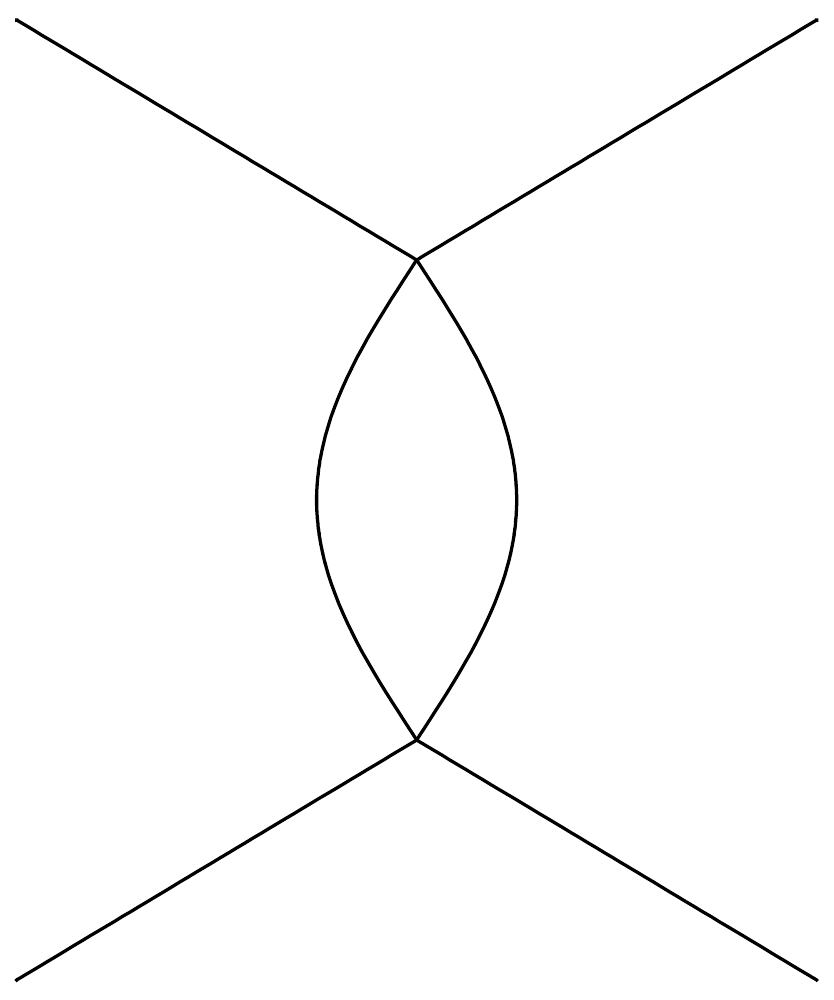}
\end{center}
\caption{One-loop scattering $2\phi\to2\phi$. Diagrams 1--9 (left block): In Jordan frame it is due to graviton exchange. Diagrams 10--12 (right block): In Einstein frame it is due to a 4-point vertex. Top row: t-channel, middle row: u-channel, bottom row: s-channel.}
\label{Scattering1lp}
\end{figure}

What about in the Einstein frame? To address this question we need to focus on the kinetic sector, which we presented earlier in eq.~(\ref{actionE}). By expanding for small $\phi$, we have the following contributions
\beq
(\partial\phi)^2+{6\,\xi^2\over\mpl^2}\phi^2(\partial\phi)^2+\ldots
\eeq
The second term appears to represent a dimension 6 interaction term with cutoff $\Lambda=\mpl/\xi$. We can compute $2\phi\to2\phi$ scattering at tree-level using this single 4-point vertex (Fig.~\ref{Scattering} diagram 4). We find the scaling $\mathcal{M}\sim\xi^2 E^2/\mpl^2$, but when the external particles are put on-shell we find that the result is zero. This is also true at arbitrary loop order (see Fig.~\ref{Scattering1lp} for the 1-loop diagrams) and for {\em any} scattering process (despite the power counting estimates of Ref.~\cite{Burgess}). The reason for this was alluded to earlier: in this single field case, we can perform a field redefinition to $\sigma$ (eq.~(\ref{kinetic})) which carries canonical kinetic energy. Hence the kinetic sector is that of a free theory, modulo its minimal coupling to gravity, giving the correct cutoff $\Lambda=\mpl$.
This field redefinition can always be done for a single field. Once the field redefinition is performed we see that the only effective interactions that will be generated, in the absence of including contributions from the potential $V$, are Planck suppressed.
Hence scattering amplitudes generated from purely the gravity and kinetic sectors allow the theory to be well defined up to Planckian energies. This shows that naive power counting can be misleading.
However, the potential $V$ causes the cutoff to be greatly reduced, as we now explain.

\subsection{The Potential}

Next we examine the potential $V$. This potential seems particularly strange in the Einstein frame, where it takes the following form
\beq
V_E(\sigma)={{\lambda\over 4}\phi(\sigma)^4\over \left(1+{\xi\phi(\sigma)^2\over\mpl^2}\right)^2}={\lambda\over 4}\sigma^4-{\lambda\,\xi^2 \over\mpl^2}\sigma^6+\ldots
\eeq
In this case it would seem that the theory breaks down for $\Lambda\sim\mpl/\xi$.
It is important to compute the quantum corrections. 
Recall the general form for the one-loop quantum corrected potential \cite{Coleman}
\beq
V_{\mbox{\tiny{1 loop}}}(\sigma)=V_{\mbox{\tiny{cl}}}(\sigma)+
{1\over64\pi^2}\left(V''_{\mbox{\tiny{cl}}}(\sigma)\right)^2\ln\!\left(V''_{\mbox{\tiny{cl}}}(\sigma)\right)+\ldots.
\eeq
Since the full potential is slowly varying it is simple to check that the corrections are always small, especially for small $\lambda$. 
Since the one-loop quantum corrections are small, there is some naive hope that this theory may make sense; this appears to be the philosophy advocated in recent studies of Higgs-inflation. 
On the other hand, one might take the point of view that by simply expanding around small $\sigma$, the theory fails when the higher order terms are comparable to the lower order terms.
The breakdown of this theory would appear in many-particle hard scattering processes, such as $3\to3$.

The behavior of a scalar field theory with a slowly varying non-polynomial potential is considered an open problem in field theory. 
However, a perturbative analysis suggests that the theory fails at $\Lambda\sim\mpl/\xi$. For instance, if we consider tree-level scattering of $2\to 2+n$ particles, then its cross-section should scale as $\sim (\lambda^2/E^2)(E^{n}\xi^{n}/\mpl^{n})$, which indicates 
that the perturbative description breaks down
for $E\gtrsim\lambda^{-2/n}\mpl/\xi\to\mpl/\xi$ for large $n$.
Note that this conclusion arises only if one studies scattering amplitudes involving vertices provided by the potential $V$.
In particular, this necessarily requires scattering amplitudes to involve powers of $\lambda$. Such factors of $\lambda$ were not included in eq.~(3.15) of Ref.~\cite{Burgess}, who focussed exclusively on the gravity and kinetic sectors. Hence, the scattering estimates provided in eq.~(3.15) of Ref.~\cite{Burgess} do not apply in the case of a singlet scalar, since they arise exclusively from the gravity and kinetic sectors alone (as we analyzed in the previous section).  However, by including corrections from the potential, the theory likely fails at $\sim\mpl/\xi$.

To say it differently, the Einstein frame Lagrangian acquires a shift symmetry $\sigma\to\sigma+\sigma_0$ in the $\lambda\to0$ limit, which protects scattering amplitudes from becoming large. This also occurs in the Jordan frame, though the symmetry is nonlinearly realized. However, finite $\lambda$ breaks the symmetry and likely compromises the theory.


\subsection{Multiple Fields}

In the case of $N$-fields, the story is rather different. Here we can just focus on the gravity and kinetic sectors to understand that the effective theory breaks down. Let us begin the discussion again in the Jordan frame. Expanding the metric around flat space as before, we have the set of dimension 5 interactions
\beq
{\xi\over\mpl}\vec\phi\cdot\vec\phi\,\square h
\eeq
Consider particles $\phi_1$ and $\phi_2$.
Lets compute the tree-level scattering process $\phi_1+\phi_2\to\phi_1+\phi_2$ due to the exchange of one graviton.
The scattering matrix scales as earlier
\beq
\mathcal{M}_{\mbox{\tiny{c}}}(\phi_1+\phi_2\to\phi_1+\phi_2)\sim{\xi^2 E^2\over\mpl^2}.
\eeq
But unlike the case of identical particles, this process only occurs through a single channel -- the t-channel (Fig.~\ref{Scattering} diagram 1). Hence there is no cancellation. This means that the theory {\em does} become strongly coupled at $\Lambda=\mpl/\xi$. Furthermore, the probability for a pair of different particles to scatter off each other grows without bound, violating unitarity (as noted in Refs.~\cite{Burgess,Atkins:2010eq}).

In the Einstein frame the kinetic sector takes the following form for small $\vec\phi$
\beq
\partial\vec\phi\cdot\partial\vec\phi+{6\,\xi^2\over\mpl^2}(\vec\phi\cdot\partial\vec\phi)^2+\ldots
\eeq
It is impossible to perform a field re-definition to bring this into canonical form.
The second term introduces cross terms between the different fields: $\xi^2\phi_1\partial\phi_1\phi_2\partial\phi_2/\mpl^2$. As in the Jordan frame, this allows for $\phi_1+\phi_2\to\phi_1+\phi_2$ scattering to take place at tree-level through a simplified vertex with $\mathcal{M}\sim \xi^2 E^2/\mpl^2$. This confirms that the theory is strongly interacting at  $\Lambda=\mpl/\xi$.

But what if we focus purely on the scattering of identical particles $2\phi_1\to2\phi_1$? Then at tree-level we seem fine, due to the cancellation among the 3 channels. However, the strong coupling between $\phi_1$ and $\phi_2$ ensures that quantum corrections from $\phi_2$ particles running in the loop are large and do not cancel among channels. In Fig.~\ref{Scattering1lp} we have included all 1PI one-loop diagrams in the Jordan frame (9 diagrams in the left block) and Einstein frame (3 diagrams in the right block). 
If only a single field $\phi_1$ is present then cancellation occurs from summing the 9 diagrams in Jordan frame, or summing the 3 diagrams in Einstein frame. When $\phi_2$ is included it only appears in the loop in the 3rd column of diagrams in the Jordan frame, and in all 3 diagrams in the Einstein frame but with a truncated vertex rule. This prevents cancellation among the diagrams. 

Note that the breakdown of the theory at $\Lambda=\mpl/\xi$ is due to scalar fields running in a loop, not gravitons. 
Hence this represents a breakdown of the matter sector, not a quantum-gravity effect per se.
In order to eliminate the ensuing divergences, we require the introduction of new operators into the classical Lagrangian:
\beq
{c_1\over\Lambda^4}(\partial\vec\phi\cdot\partial\vec\phi)^2\,\,,\,\,\,\,\,\,\,\,\,\,
{c_2\over\Lambda^4}(\partial_\mu\vec\phi\cdot\partial_\nu\vec\phi)(\partial^\mu\vec\phi\cdot\partial^\nu\vec\phi),
\eeq
with $\Lambda=\mpl/\xi$ and the $c_i$ cannot be much smaller than $\mathcal{O}(4\pi)^{-2}$ without fine tuning.
In addition, the quantum field theory will also generate corrections to the potential, such as $\bar{c}_1\lambda^2(\vec\phi\cdot\vec\phi)^3/\Lambda^2$ and $\bar{c}_2\lambda^2(\vec\phi\cdot\vec\phi)^4/\Lambda^4$. Note the factor of $\lambda^2$ here, which arises because the potential is protected by a shift symmetry in the $\lambda\to 0$ limit.

How do such operators affect the inflationary model? Lets focus on the end of inflation where $\phi\sim\mpl/\sqrt{\xi}$. Here the kinetic and potential terms in the classical Lagrangian are comparable, with value $V\sim(\partial\phi)^2\sim\lambda\,\mpl^4/\xi^2$. The new dimension 8 operators would be $\sim c\,\lambda^2\mpl^4$; parametrically larger than the included operators by a factor $c\,\lambda\,\xi^2$, destroying the theory for $\xi\gg 1$. This is relevant to the Standard Model Higgs which is comprised of multiple scalars.
It should be pointed out that the above reasoning is based on a perturbative analysis around $\langle\phi\rangle=0$, which is obviously not true during inflation, where $\langle\phi\rangle\sim\sqrt{N_e}\mpl/\sqrt{\xi}$. However, the moment inflation ends, the field will oscillate around $\langle\phi\rangle=0$ and the above analysis is essentially correct, i.e., the multiple degrees of freedom will interact strongly.  This means, unfortunately, that one cannot describe the reheating era using this low energy effective field theory, or connect inflation to low energy Higgs physics.

\section{An Analogy - the Pion Lagrangian}\label{Pion}

If this distinction between the behavior of a single scalar field theory compared to a multiple scalar field theory seems strange, let us recall a familiar case where the same is true.

Consider the Lagrangian of the $\sigma$-model
\beq
\mathcal{L}=-{1\over2}\partial\phi_n\partial\phi_n+{\mu^2\over2}\phi_n\phi_n-{\lambda\over4}(\phi_n\phi_n)^2
\eeq
where $n$ is summed over $n=1,2,\ldots,N,N+1$, with the first $N$ components forming a vector $\vec\phi$ 
and the last component $\phi_{N+1}$ a scalar.
After spontaneous symmetry breaking, it is useful to use fields that make it manifest that the potential only depends on the length of $\phi_n$, which we call $\sigma(x)\equiv\sqrt{\sum_n\phi_n(x)^2}$, and the Goldstone bosons, which we call $\vec\zeta(x)$. Following Ref.~\cite{Weinberg} the original Lagrangian can be recast as
\beq
\mathcal{L}=-{1\over2}(\partial\sigma)^2-2\sigma^2{\partial\vec\zeta\cdot\partial\vec\zeta\over(1+\vec\zeta\cdot\vec\zeta)^2}
+{\mu^2\over2}\sigma^2-{\lambda\over4}\sigma^4.
\eeq
At low energies, the massive field $\sigma$ will relax to its minimum at $\langle\sigma\rangle=\mu/\sqrt\lambda$.
By defining $F\equiv2\langle\sigma\rangle$ and introducing properly normalized Goldstone bosons $\vec\pi(x)\equiv F\,\vec\zeta(x)$, the low energy Lagrangian for the Goldstone bosons is
\beq
\mathcal{L}=-{1\over2}{\partial\vec\pi\cdot\partial\vec\pi\over(1+\vec\pi\cdot\vec\pi/F^2)^2}.
\label{pionLag}
\eeq
For 3 pions it is well known that this theory breaks down for energies much larger than $F$, which can be confirmed by computing scattering amplitudes. A rough estimate for the breakdown of the quantum theory is $\Lambda=4\pi F$, where the $4\pi$ comes from loop integrals.

The $2\to2$ tree-level scattering process has the matrix element
\bea
&&\mathcal{M}(\pi_a+\pi_b\to\pi_c+\pi_d) \nonumber \\
&&=4F^{-2}\left[\delta_{ab}\delta_{cd}(-p_ap_b-p_cp_d)+\delta_{ac}\delta_{bd}(p_ap_c+p_bp_d)+\delta_{ad}\delta_{bc}(p_ap_d+p_bp_c)\right]
\eea
which is non-zero whenever some of the $a,b,c,d$ are different (scaling as $\sim E^2/F^2$), but is zero if $a=b=c=d$ and the $\pi$'s are put on-shell ($p^2=0$). The latter is the case for a single field. The same is true for loop corrections. As earlier, what is special about the $N=1$ case is that the Lagrangian (\ref{pionLag}) is really a free field theory in disguise (which can be made manifest through a field redefinition), but it is unambiguously an interacting field theory for $N>1$. 

On the other hand, even for $N=1$, if we explicitly break the global symmetry, then the (pseudo)-Goldstone bosons acquire a mass and a potential $V$. This potential is non-polynomial and likely causes the theory to fail for $E\gg F$.

\section{Curvature-Squared Models}\label{Curvature}

Let us now turn to another type of non-standard model for inflation, which comes from the inclusion of curvature-squared terms in the action \cite{Starobinsky}
\beq
S=\int d^4 x \sqrt{-g} \left[{1\over2}\mpl^2\mathcal{R}+\zeta\,\mathcal{R}^2\right],
\eeq
with $\zeta=\mathcal{O}(10^8)$ to achieve the correct amplitude of density fluctuations. Although we consider this model to be ad hoc
and not necessarily ``natural", it is still worthwhile to discuss whether this theory makes sensible inflationary predictions or not.

Since $\zeta\gg 1$ we again should ask if scattering amplitudes become large at energies well below $\mpl$.
In particular, by expanding around flat space, the curvature-squared term introduces the following dimension 6 and 7 operators into the action
\beq
{\zeta\over\mpl^2} (\square h)^2,\,\,\,\,\,\,\,\,\,\,\,
{\zeta\over\mpl^3} (\square h)^2 h,
\eeq
respectively. The former modifies the graviton propagator at high energies.
The latter introduces a 3-point vertex, which can be used to construct tree-level graviton-graviton scattering processes.
In particular, each of the t, u, and s-channels scale to leading order as
\beq
\mathcal{M}_c(2\,h_{\mu\nu}\to2\,h_{\mu\nu})\sim{\zeta^2 E^6\over\mpl^6},
\eeq
suggesting that the theory violates unitarity above the cutoff $\Lambda=\mpl/\zeta^{1/3}$, as claimed recently in \cite{Burgess} and criticized on that basis. However, as was the case in Section \ref{Single}, if we sum over all diagrams, then this piece vanishes.
Instead, the leading non-zero piece scales as $\mathcal{M}_{\mbox{\tiny{tot}}}\sim (E/\mpl)^2$.
(Note that there is also a 4-point graviton vertex and that the correction to the graviton propagator is important here).
So according to the tree-level analysis of $2\to2$ scattering, the correct cutoff is $\Lambda=\mpl$.

The most transparent way to analyze the perturbative quantum corrections to this theory is in the frame in which the degrees of freedom are manifest and to use fields with diagonal kinetic sectors.
Since the Jordan frame contains higher derivatives, due to the $\mathcal{R}^2$ term, we choose to switch to the Einstein frame to make the degrees of freedom manifest.
Ref.~\cite{Burgess} claimed that this theory fails equally in any frame, including the Einstein frame. 
So lets re-write the action as a scalar-tensor theory. This involves a canonical gravity sector, a canonical kinetic sector for some scalar $\sigma$, and a potential $V_E(\sigma)$ which exhibits $\mathcal{O}(1)$ variations on $\Delta\sigma\sim\mpl$. It can be shown that this gives
\beq
S=\int d^4 x\sqrt{-g_E}\left[{1\over2}\mpl^2\mathcal{R}_E-{1\over2}(\partial\sigma)^2-{\mpl^4\over 16\,\zeta}
\left(1-\exp(-\sqrt{2/3}\,\sigma/\mpl)\right)^2   \right].
\label{scalartensor}\eeq
This action seems rather innocuous, but according to Ref.~\cite{Burgess} this theory still has a cutoff at $\Lambda=\mpl/\zeta^{1/3}$.
We begin by explaining how one could arrive at this conclusion by a naive power counting estimate. We then explain why this is incorrect and why the correct answer is $\Lambda=\mpl$.

Let us expand the potential about its minimum at $\sigma=0$: 
$V_E(\sigma)={\mpl^2\over24\zeta}\sigma^2+\ldots$
We see that the field $\sigma$ has a mass-squared of $m_\sigma^2=\mpl^2/(12\,\zeta)$.
This non-zero mass will obviously play an important role in computing low energy scattering amplitudes.
For instance, consider the 1-loop scattering process 
involving $2\to2$ graviton scattering due to $\sigma$ running in a loop. We use 4 insertions of the following 3-point interaction term from the kinetic sector
\beq
\sim {h\over\mpl}(\partial\sigma)^2.
\eeq
This gives rise to the following contribution to the matrix element
\beq
\mathcal{M}(2\,h_{\mu\nu}\to2\,h_{\mu\nu})\sim{1\over\mpl^4} \int d^4q{q_1\!\cdot\!q_2\,\,q_2\!\cdot\!q_3\,\,q_3\!\cdot\!q_4\,\,q_4\!\cdot\!q_1\over (q_1^2-m_\sigma^2)(q_2^2-m_\sigma^2)(q_3^2-m_\sigma^2)(q_4^2-m_\sigma^2)}.
\eeq
This loop integral can be estimated using, for instance, dimensional regularization. Its scaling depends critically on the energy scale.
For input energy $E\ll m_\sigma$, the denominator can be approximated as $m_\sigma^8$. If one were to then approximate the numerator with the high energy behavior, the finite part would be estimated as
\beq
\mathcal{M}(2\,h_{\mu\nu}\to2\,h_{\mu\nu})\sim {E^{12}\over\mpl^4m_\sigma^8}\sim{\gamma^4E^{12}\over\mpl^{12}}.
\eeq
If this scaling were correct up to arbitrarily high energies, then indeed the theory would have a cutoff at $\Lambda=\mpl/\zeta^{1/3}$. 
However, this scaling is invalid at high energies. Instead, for $E\gg m_\sigma$, the matrix element scales as
\beq
\mathcal{M}(2\,h_{\mu\nu}\to2\,h_{\mu\nu})\sim {E^{4}\over\mpl^4},
\eeq
indicating that the correct cutoff is $\Lambda=\mpl$. A similar argument for the cutoff goes through for {\em any} scattering process, including tree-level or arbitrary loop level.

Hence, if low energy scattering amplitudes are taken at face value, they naively suggest that the theory has a cutoff at $\mpl/\zeta^{1/3}$. However, a full calculation at the appropriate energy scale reveals that the correct cutoff of the scalar-tensor theory defined by (\ref{scalartensor}) is $\mpl$. 
This is because the potential varies on scales $\Delta\sigma\sim\mpl$.
This means that the inflationary predictions make sense. 



\section{Conclusions}\label{Conclusion}

Treated classically, a singlet scalar $\phi$ with non-minimal coupling to gravity $\xi\phi^2\mathcal{R}$ and $\xi\gg 1$
 can drive a phase of slow-roll inflation. 
By examining the gravity and kinetic sectors alone, as in Ref.~\cite{Burgess},
it is tempting to declare that such theories become strongly interacting at $\mpl/\xi$, which would spoil the model. Such a conclusion arises from power counting estimates of scattering processes in the Jordan frame without summing all diagrams to check for any possible cancellations. Here we have shown that cancellations do occur in the single field case, giving a Planckian cutoff. 
However, the potential sector is problematic: the Einstein frame potential is non-polynomial, varying on scales $\Delta\sigma\sim\mpl/\xi$. Although its Coleman-Weinberg quantum corrections \cite{Coleman} are small and the behavior of such a theory is considered an open problem in field theory, it is doubtful if high energy scattering amplitudes could be well behaved and if there is any sensible UV completion of the theory. 
This suggests that the true cutoff is $\sim\mpl/\xi$ due to the potential sector.

On the other hand, these power counting estimates in the gravity and kinetic sectors {\em do} apply in the multi-field case: a vector of fields $\vec\phi$ is problematic as $\phi_a+\phi_b\to\phi_c+\phi_d$ scattering becomes strong at energies $E\gtrsim\mpl/\xi$ relevant to inflation (as noted in \cite{Burgess,Atkins:2010eq}) and large quantum corrections are generated.
This requires new physics to intervene or that something peculiar happens in the RG flow that rescues the theory, such as flow to a UV fixed point. But the latter scenario seems rather unlikely. Our conclusions were shown to be true in both the Jordan and Einstein frames.

So if the Standard Model Higgs were a gauge singlet, then Higgs-inflation would be well behaved in the pure gravity and kinetic sectors, requiring a detailed analysis of the potential sector to draw conclusions about the quantum theory.
However, since it is comprised of 4 real scalars in a complex doublet, it falls into the multi-field category and fails even at the level of tree-level scattering due to graviton exchange; breaking down due to Goldstone bosons. Explicitly
identifying this difference between the single field case and the multi-field case
was previously missed by us \cite{DeSimone:2008ei} and others. 
Similarly, this may spoil other attempts to embed inflation into particle physics models with multiple scalar fields \cite{Hertzberg:2007wc}, unless the underlying UV theory carries an appropriate structure. 

Our conclusions were drawn from an expansion around $\langle\phi\rangle=0$. Although one might be concerned that this does not reflect the inflationary regime where $\langle\phi\rangle$ is large, it does accurately reflect the end of inflation (the reheating era) in which $\phi$ oscillates around $\phi=0$. Hence our conclusions are quite robust in demonstrating that the reheating era cannot be described within effective field theory, i.e.,  that the effective Lagrangian must have large corrections as we {\em approach} inflationary energy densities. Thus altering the inflationary Lagrangian itself.

We also examined curvature-squared models $\zeta\,\mathcal{R}^2$ \cite{Starobinsky} with $\zeta\gg 1$ (although we consider such models to be ad hoc). 
They were recently criticized in Ref.~\cite{Burgess} for failing to even make sense at scales $E\gtrsim\mpl/\zeta^{1/3}$.
Again there exists cancellation among tree-level diagrams, suggesting a Planckian cutoff. Furthermore, by formulating the quantum theory as a scalar-tensor theory in the Einstein frame (which makes the degrees of freedom manifest) we showed that the inflationary theory is well behaved. In this case, all scattering processes, whether they arise from the gravity or kinetic or potential sectors, satisfy unitarity bounds and generate very small quantum corrections for energies below $\mpl$. The reason for this is that we have a singlet model with potential that varies on scales $\Delta\sigma\sim\mpl$ (as opposed to non-minimally coupled models which vary on scales $\Delta\sigma\sim\mpl/\xi$.)

\begin{acknowledgments}
We would like to thank R.~Jackiw, D.~Kaiser, R.~Kallosh, K.~Lee, A.~Linde, M.~Tegmark, A.~Westphal, F.~Wilczek, S.~Yaida, and especially A.~De Simone for useful discussions.
This work is supported in part by the U.S. Department of Energy (DoE) under contract No. DE-FG02-05ER41360. 
\end{acknowledgments}

\section*{Note Added}
 As this work was completed an interesting paper came out one day earlier \cite{Burgess:2010zq}, with similar results.
 These authors show explicitly that the cutoff at $\Lambda=\mpl/\xi$ for the Standard Model Higgs appears independent of the choice of gauge, such as covariant gauge and unitary gauge.


\end{document}